\documentclass[draft]{agujournal}
\draftfalse

\usepackage{graphics,graphicx}
\usepackage{graphicx,epstopdf}
\usepackage{amssymb}
\usepackage{amsfonts}
\usepackage{esint} 
\usepackage{braket}

\journalname{Journal of Geophysical Research-Space Physics}

\begin{document}

%
%


\title{Solar wind interaction with the Martian upper atmosphere: Roles of the cold thermosphere and hot oxygen corona}

%
%




\authors{Chuanfei Dong,\affil{1,2} Stephen W. Bougher,\affil{3} Yingjuan Ma,\affil{4} Yuni Lee,\affil{5} Gabor Toth,\affil{3} Andrew F. Nagy,\affil{3} Xiaohua Fang,\affil{6} Janet Luhmann,\affil{7} Michael W. Liemohn,\affil{3} Jasper S. Halekas,\affil{8} Valeriy Tenishev,\affil{3} David J. Pawlowski,\affil{9} Michael R. Combi\affil{3}}

\affiliation{1}{Department of Astrophysical Sciences, Princeton University, Princeton, New Jersey, USA}

\affiliation{2}{Princeton Center for Heliophysics, Princeton Plasma Physics Laboratory, Princeton University, Princeton, New Jersey, USA}

\affiliation{3}{Department of Climate and Space Sciences and Engineering, University of Michigan, Ann Arbor, Michigan, USA.}

\affiliation{4}{Department of Earth and Space Sciences, University of California, Los Angeles, California, USA.}

\affiliation{5}{NASA Goddard Space Flight Center, Greenbelt, Maryland, USA.}

\affiliation{6}{Laboratory for Atmospheric and Space Physics, University of Colorado, Boulder, Colorado, USA.}

\affiliation{7}{Space Sciences Laboratory, University of California, Berkeley, California, USA.}

\affiliation{8}{Department of Physics and Astronomy, University of Iowa, Iowa City, Iowa, USA.}

\affiliation{9}{Department of Physics and Astronomy, Eastern Michigan University, Ypsilanti, Michigan, USA.}






\correspondingauthor{Chuanfei Dong}{dcfy@princeton.edu}




\begin{keypoints}
\item Despite the similar ion loss rate calculated from 1D and 3D atmospheres, the latter are required to adequately reproduce MAVEN observations.
\item The hot oxygen corona plays an important role in protecting the Martian ionosphere and thermosphere from the solar wind erosion.
\item The thermospheric oxygen atom is the primary neutral source for O$^+$ ion escape during the relatively weak solar cycle 24.
\end{keypoints}

%
%


\begin{abstract}
We study roles of the thermosphere and exosphere on the Martian ionospheric structure and ion escape rates in the process of the solar wind-Mars interaction. We employ a four-species multifluid MHD (MF-MHD) model to simulate the Martian ionosphere and magnetosphere. The \emph{cold} thermosphere background is taken from the Mars Global Ionosphere Thermosphere Model (M-GITM) and the {hot} oxygen exosphere is adopted from the Mars exosphere Monte Carlo model - Adaptive Mesh Particle Simulator (AMPS). A total of four cases with the combination of 1D (globally averaged) and 3D thermospheres and exospheres are studied.

The ion escape rates calculated by adopting 1D and 3D atmospheres are similar; however, the latter are required to adequately reproduce MAVEN ionospheric observations. In addition, our simulations show that the 3D hot oxygen corona plays an important role in preventing planetary molecular ions (O$_2^+$ and CO$_2^+$) escaping from Mars, mainly resulting from the mass loading of the high-altitude exospheric O$^+$ ions. The \emph{cold} thermospheric oxygen atom, however, is demonstrated to be the primary neutral source for O$^+$ ion escape during the relatively weak solar cycle 24.
\end{abstract}

%
%

%


%
%
%
%

\section{Introduction}\label{sec:Intro}

Unlike Earth and Venus, Mars with a relatively weak surface gravity allows an extended corona of hot oxygen that can partially escape to space \citep{Wallis1978,Ip1988,Nagy1988,Fox1993}. Being the most important reaction, the dissociative recombination of O$_2^+$ (deep in the dayside thermosphere/ionosphere) is responsible for producing most of dayside exospheric hot atomic oxygen (O$_2^+$ + e $\longrightarrow$ O$^*$ + O$^*$), therefore, the distribution of the hot oxygen exosphere is asymmetric around the Mars globe \citep{Valeille2009,Lee2015a}. Note that it is not only the model that shows the hot O distribution is asymmetric, but it is also seen in the MAVEN Imaging Ultraviolet Spectrograph (IUVS) data \citep[e.g.,][]{Lee2015b,Leblanc2017}. In addition to the dissociative recombination of O$_2^+$, the sputtering caused by collisions between the pickup ions (e.g., O$^+$) and the Martian thermospheric background may also be an important source for the hot corona \citep{Luhmann1991,Johnson1998,Leblanc2018}. Compared with the \emph{cold} thermospheric background, the \emph{hot} (or energetic) oxygen has a thermal speed, $\braket{v_O}=(2k_BT_O/m_O)^{\frac{1}{2}}$, higher than the local thermal speed of the thermosphere.

Besides the strong day-night asymmetry exhibited in the hot oxygen density distribution, the major neutral species in the Martian thermosphere (CO$_2$ and O) are also distributed asymmetrically about the planet \citep[e.g.,][]{Bougher2008,Bougher2015a}. Specifically, more neutral CO$_2$ molecules are in the dayside thermosphere than on the nightside at a given altitude because the CO$_2$ global distribution is mainly controlled by the global temperature instead of the dynamics. Therefore, the thermospheric CO$_2$ density increases (decreases) on the dayside (nightside) where temperatures are higher (lower). The density distribution of atomic O (especially on the nightside), however, is mainly controlled by the day-night transport due to its relatively low mass; photochemistry may make certain contribution on the dayside oxygen density distribution. For atomic O, transport begins to have an effect as the thermospheric winds increase with increasing altitude above the region where dayside O is produced photochemically; the day-to-night atomic O distribution is impacted strongly by winds roughly above $\sim$130 km \citep{Bougher2015a}. The neutral wind can transport atomic O from dayside to nightside, resulting in a bulge of neutral O in the nightside thermosphere \citep{Bougher2015a}.

In order to capture the asymmetry in the Martian thermosphere and exosphere, three-dimensional ``whole atmosphere'' (from the ground to the exobase, 0 to $\sim$ 250 km) \citep{Bougher2015a} and exosphere \citep{Lee2015a} models are ultimately required to capture these asymmetric features. The modeled thermosphere and exosphere can be further input into a global plasma code as the neutral background, such that roles of the 3D thermosphere and exosphere on the Martian ionospheric structure and ion escape processes can be investigated in detail. Note that the incident solar wind at Mars encounters an extended hot exosphere, a conductive ionosphere, and highly localized crustal magnetic fields (the strongest of which in the southern hemisphere \citep{Acuna1999}), resulting in a complex obstacle to the solar wind that varies on all spatial and temporal scales. Among all the objects in the solar system, Mars, therefore, offers a uniquely challenging set of conditions to simulate. 

In recent years, investigations of the Martian thermosphere/ionosphere structure \citep[e.g.,][]{Withers2015,Bougher2015b}, magnetic topology \citep[e.g.,][]{Luhmann2015,Xu2016,Liemohn2017,DiBraccio2018}, and atmospheric escape rates \citep[e.g.,][]{Halekas2016,Fang2017,Egan2018} have become increasingly important because they are closely related to the evolution of the Martian atmosphere and can affect its climate over the past four billion years \citep[e.g.,][and the references therein]{Jakosky2015a,Bougher2015c,Lillis2015,Mansfield2018,Dong2018b,Jakosky2018}. In-situ spacecraft measurements \citep[e.g.,][]{Lundin2013,Ramstad2015,Dong2015c,Brain2015} have greatly improved our estimates of global ion loss rates at the current epoch. By using Mars Express (MEX) Analyzer of Space Plasmas and Energetic Atoms 3 (ASPERA-3) data from June 2007 to January 2013, \citet{Lundin2013} reported that the average heavy ion escape rate increased approximately by a factor of 10, from 1 $\times$ 10$^{24}$ s$^{-1}$ (solar minimum) to 1 $\times$ 10$^{25}$ s$^{-1}$ (solar maximum). More recently, \citet{Brain2015} analyzed four months of Mars Atmosphere and Volatile EvolutioN (MAVEN) spacecraft data and estimated a net ion escape rate of $\sim$2.5 $\times$ 10$^{24}$ s$^{-1}$ by choosing a spherical shell at $\sim$1000 km above the planet with energies $>$25 eV during solar cycle maximum conditions. In addition, \citet{Liemohn2014} and \citet{Dong2015c} confirmed the substantial plume-like distribution of escaping ions from the Martian atmosphere in MEX and MAVEN observations, organized by the upstream solar wind convection electric field. It is also worth noting that the total ion loss rate increased by more than one order of magnitude during an interplanetary corona mass ejection (ICME) event observed by MAVEN on March 8th, 2015 \citep{Jakosky2015b,Dong2015b,Curry2015b,Ma2017,Luhmann2017}. Moreover, \citet{Lingam2018} found that the solar energetic protons (SEPs) associated with extreme space weather events with energies $\gtrsim$ 150 MeV can reach the Martian surface; the same cutoff value has also been presented by the Mars Science Laboratory's Curiosity rover group \citep{Hassler2014}.

In order to study the solar wind interaction with the Martian upper atmosphere, various plasma fluid models and kinetic particle codes have been developed.  A few notable examples include the multi-species single-fluid MHD models \citep{Ma2004,Ma2014}, the multifluid MHD models \citep{Harnett2006,Najib2011,Riousset2013,Riousset2014,Dong2014a,Dong2015a}, the test-particle approach \citep{Fang2008,Fang2010a,Curry2014,Curry2015a} and the hybrid particle-in-cell (hybrid-PIC) codes \citep{Modolo2016,Brecht2016}. These codes have been used to help quantify the ion escape rates from the Martian upper atmosphere through the solar wind-Mars interaction. Most of these studies can reach a reasonable agreement with the spacecraft observations. However, until now no systematic study was focused on the influence of 3D thermospheres and exospheres on the Martian ionospheric structure and ion escape rates.

\begin{figure*}[!htb]
\centering
\includegraphics[width=25pc,angle=0]{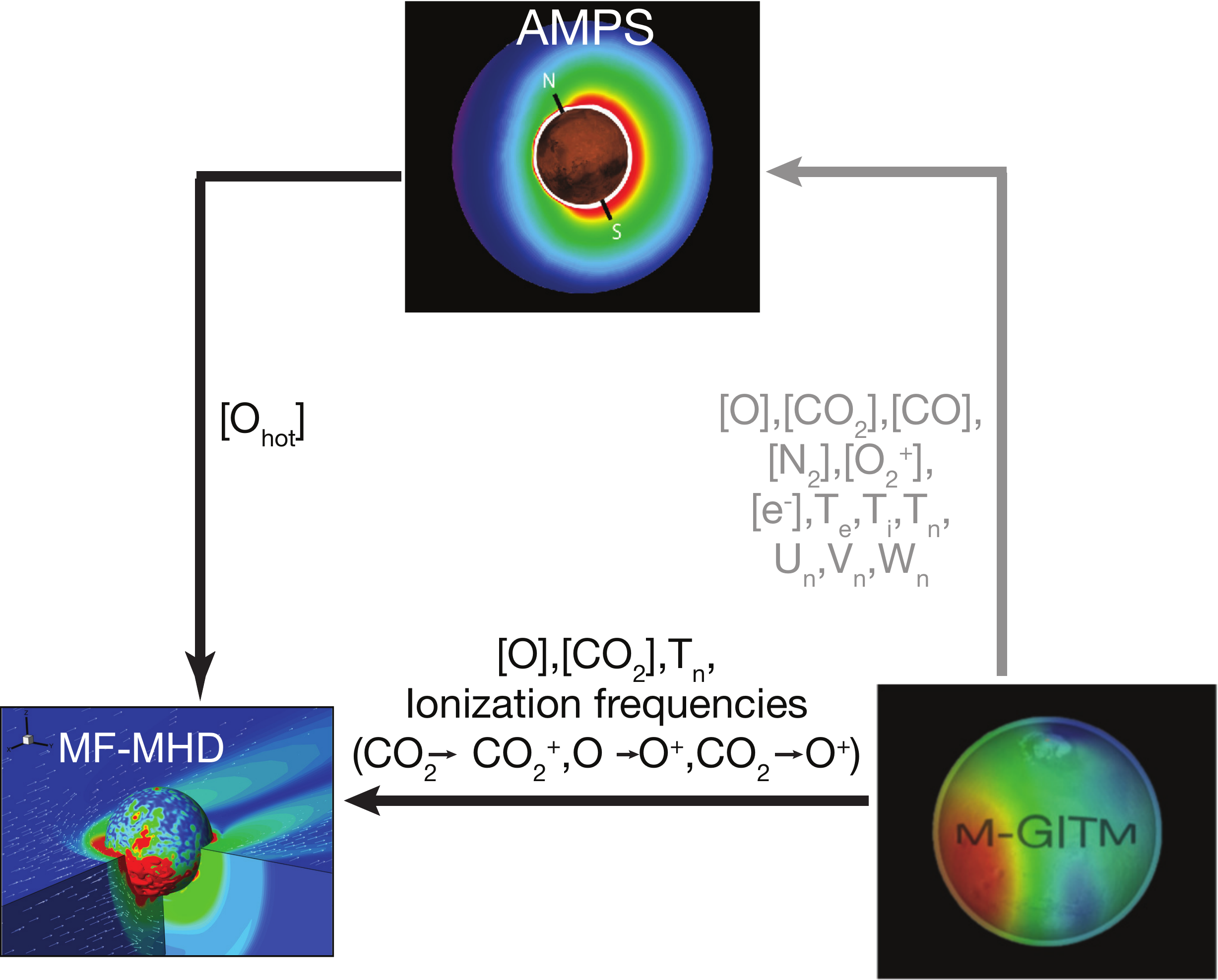}
\caption{Sketch of a one-way coupling approach between M-GITM, Mars AMPS and MF-MHD (after Figure 1 of \citet{Dong2015a}). The notation T$_n$ denotes the neutral atmospheric temperature. The quantities [O], [CO$_2$], and [O$_{hot}$] are the thermospheric \emph{cold} O, CO$_2$ and exospheric \emph{hot} O number densities, respectively. In this study, we adopt the one-way coupling indicated by the solid black lines. For the detailed study of the one-way coupling between M-GITM and Mars AMPS (solid gray line), please refer to \citet{Lee2015a}.}
\label{coupling}
\end{figure*}

In this paper, we adopt the 3D Mars thermosphere (i.e., neutral temperatures T$_n$, neutral densities n$_O$, n$_{CO_2}$, and photoionization frequencies I$_O$, I$_{CO_2}$) from the Mars Global Ionosphere Thermosphere Model (M-GITM) \citep{Bougher2015a} and the hot atomic oxygen density, n$_{O_{hot}}$, from the Mars exosphere Monte Carlo model - Adaptive Mesh Particle Simulator (AMPS) \citep{Lee2015a}. M-GITM and Mars AMPS are one-way coupled with the 3D Block-Adaptive-Tree-Solarwind-Roe-Upwind-Scheme (BATS-R-US) Mars multifluid MHD (MF-MHD) model \citep{Najib2011,Dong2014a,Dong2015a} (see Figure \ref{coupling} for the one-way coupled framework). The Mars AMPS hot oxygen corona is calculated based on the thermosphere/ionosphere background from M-GITM \citep{Lee2015a}. In the present work, the simulations are carried out for four selected cases with the combination of 1D and 3D neutral atmospheres. 

The remainder of this paper is divided into three sections. In Section \ref{modelintro}, we briefly introduce the three models employed in this study. In Section \ref{thermosphere}, we investigate the role of the 3D thermosphere on the Martian ionospheric structure and ion escape rates by means of data-model and model-model comparisons. In Section \ref{exosphere}, we study the effect of the 3D exosphere on the ion escape rate and the corresponding molecular to atomic escaping ion ratio (O$_2^+$+CO$_2^+$)/O$^+$ through model-model comparisons. Discussion and Conclusions are summarized in the last section. 

\section{Model Description} \label{modelintro}

In this section, M-GITM, AMPS, and MF-MHD are briefly introduced. All these models have been adopted to support the MAVEN mission activities (2014-2018).

\subsection{Mars Global Ionosphere Thermosphere Model (M-GITM)} \label{M-GITMintro}

Mars Global Ionosphere Thermosphere Model (M-GITM) \citep{Bougher2015a}, combines the terrestrial GITM framework \citep{Ridley2006,Deng2008} with the fundamental physical parameters, ion-neutral chemistry, and key radiative processes for Mars in order to capture the basic observed features of the thermal, compositional, and dynamical structure of the Mars atmosphere from the ground to the exobase (0 -- 250 km). M-GITM currently solves for three-dimensional neutral and ion densities, as well as neutral temperatures and winds around the globe. Key neutral species presently include: CO$_2$, CO, O, N$_2$, O$_2$, Ar, and He. Five key photochemical ion species currently include: O$^{+}$, O$_{2}^{+}$, CO$_{2}^{+}$, N$_{2}^{+}$ and NO$^{+}$. Typically, production runs are conducted for a 5 $\times$ 5 degree longitude-latitude grid, with a constant 2.5 km vertical resolution ($\sim$ 0.25 scale height, $H_s=k_BT/mg$, where $k_B$ is Boltzmann constant, $g$ is the acceleration due to planetary gravity, $T$ is the neutral temperature, and $m$ is the mass of the neutral species).

\begin{figure*}[!htb]
\centering
\includegraphics[width=30pc,angle=0]{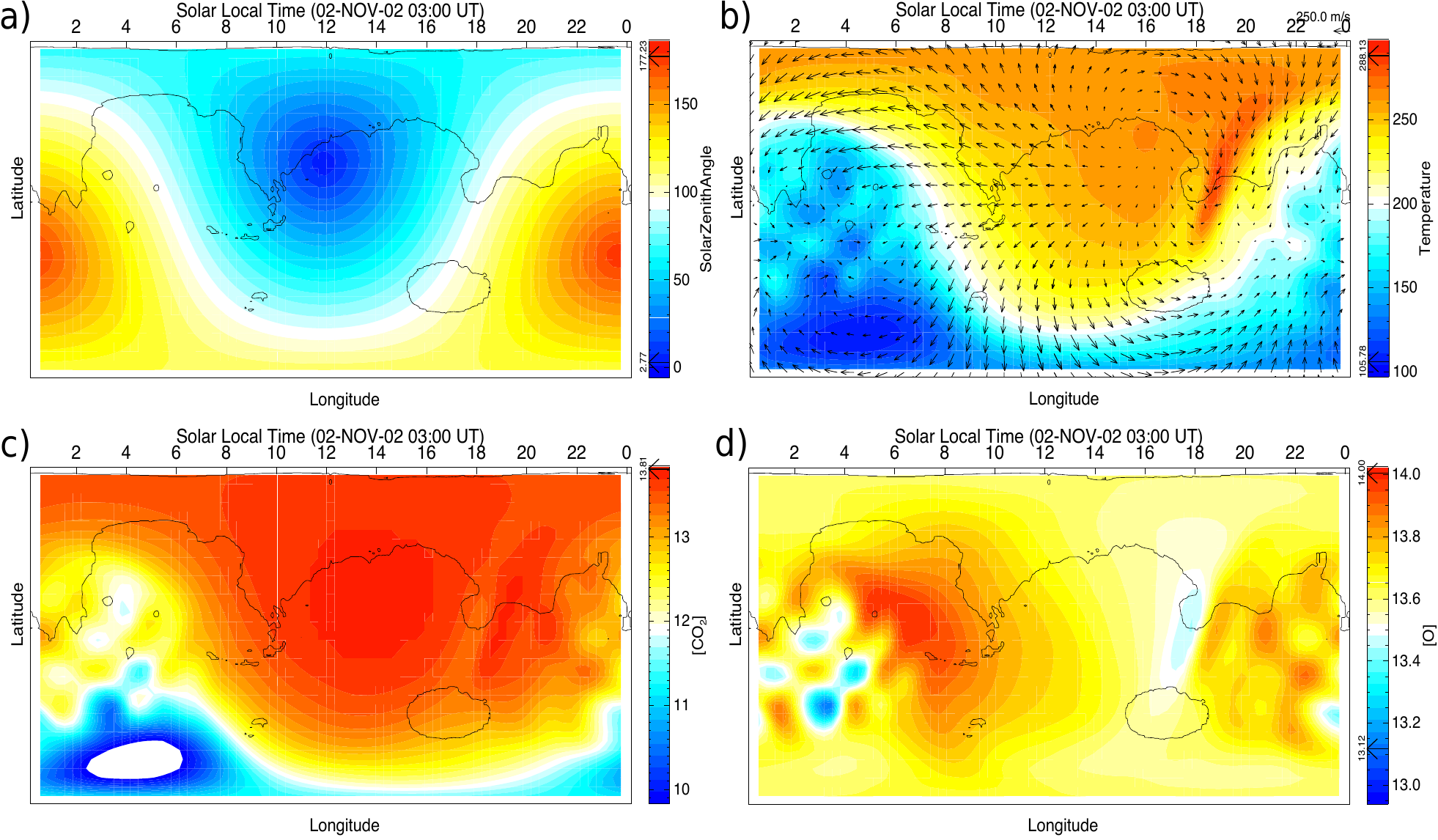}
\caption{The color contours of (a) solar zenith angle (SZA, in degree), (b) Temperature (in K), (c) log10 CO$_2$ densities (in m$^{-3}$), and (d) log10 atomic O densities (in m$^{-3}$) at $\sim$ 200 km (exobase) altitude for aphelion solar moderate conditions (APHMOD, Ls=90, F$_{10.7}$=130). The arrows in Figure \ref{MGITM} (b) indicate the relative magnitude and the direction of the horizontal winds. All the vertical axes (i.e., latitude) range from $-$90$^{\circ}$ to 90$^{\circ}$. The white color highlights the regions below the low saturation of the colorbar.}
\label{MGITM}
\end{figure*}

M-GITM validation studies thus far have focused upon simulations for a range of solar cycles and seasonal conditions \citep{Bougher2015a,Bougher2015b,Bougher2017}. Figure \ref{MGITM}a shows the solar zenith angle (SZA) distribution around Mars' globe for aphelion solar moderate conditions (APHMOD) in the Geographic (GEO) coordinate system. The subsolar point (i.e., where SZA=0) is located in the northern hemisphere. An inspection of Figure \ref{MGITM}b reveals that solar-driven exobase temperatures peak in the middle afternoon at the subsolar latitude (25$^{\circ}$N). The warmer temperature near the evening terminator (LT = 18) is a result of the dynamical heating due to the convergent zonal winds \citep{Bougher2015a}. The asymmetric distribution of CO$_2$ in latitude (Figure \ref{MGITM}c) is closely related to the asymmetric diurnal temperature distribution (Figure \ref{MGITM}b). Conversely, Figure \ref{MGITM}d presents atomic oxygen density distributions for which dayside-produced O is transported to the nightside by the thermospheric wind system, where it subsequently accumulates at low-to-middle latitudes around LT = 4--8. All the features shown in Figure \ref{MGITM} indicate the importance of adopting the 3D M-GITM thermosphere in a global plasma model in order to reproduce the ionospheric structure and accurately estimate the ion escape rates in the process of the solar wind-Mars interaction. 

\subsection{Mars Adaptive Mesh Particle Simulator (AMPS)}\label{M-AMPSintro}

The code we employ to model the Martian exosphere is the 3D Mars Adaptive Mesh Particle Simulator (AMPS), which runs in the test-particle mode using the Direct Simulation Monte Carlo (DSMC) method \citep{Bird1994}. The AMPS code \citep{Tenishev2008,Tenishev2013,Lee2015a} is a well-tested code for a wide range of kinetic problems in rarefied gas regime. Examples of the AMPS applications include the cometary coma and the exospheres of Mars, Mercury and the Moon. The 3D structure and photochemical loss of hot oxygen particles from Mars have been investigated by taking advantage of the one-way coupled framework between Mars AMPS and M-GITM \citep{Lee2015a,Lee2015b}. As shown in Figure \ref{coupling}, AMPS calculates the hot atomic oxygen density distribution based on the thermospheric background (i.e., neutral species O, CO$_2$, N$_2$, CO) from M-GITM. 

Compared with the previous version where it assumed idealized hard sphere collisions and only isotropic scattering in the center of mass frame \citep{Valeille2009}, the current AMPS considers a more realistic description for the collisions between hot O and ambient species by adopting a forward scattering collision scheme with the angular differential scattering cross sections from \citet{Kharchenko2000}. The related integrated cross sections (in cm$^{2}$) are 1.2$\times$10$^{-14}$ for O-CO$_2$, 6.4$\times$10$^{-15}$ for O-O, and 1.8$\times$10$^{-14}$ for both O-N$_2$ and O-CO. The current AMPS (by adopting the forward scattering scheme) produces a more intensive (and closer to observed) hot oxygen corona than the previous case by adopting the isotropic scattering scheme, and thus enhances the corresponding photochemical escape rate \citep{Lee2015a}. However, there still exists certain discrepancy between MAVEN observations and AMPS predictions \citep{Lee2015a}; further improvement of  model predications of hot oxygen corona is an ongoing MAVEN effort. 

The motion of each hot particle is influenced by the gravitational field of the Mars and modified by collisions with the background thermospheric species. The collision in the code depends on the rate of change in the background densities (i.e., rate of change in collision frequency). Although the nominal cell size is about 60 km at the model lower boundary (at 100 km altitude above the Martian surface), the large grid size does not prevent AMPS from capturing the variation in the Martian ionosphere and thermosphere. In AMPS, each macro-particle is initialized based on the thermospheric background prescribed by M-GITM (stored in an additional data table) at its resolution. It is noteworthy that the hybrid-PIC codes \citep[e.g.,][]{Modolo2016,Brecht2016} have similar grid resolution for studying the solar-wind Mars interaction. The AMPS computational domain extends to 6 Mars radius (one Mars radius, R$_M\sim$ 3396 km).

\begin{figure*}[!htb]
\centering
\includegraphics[width=30pc,angle=0]{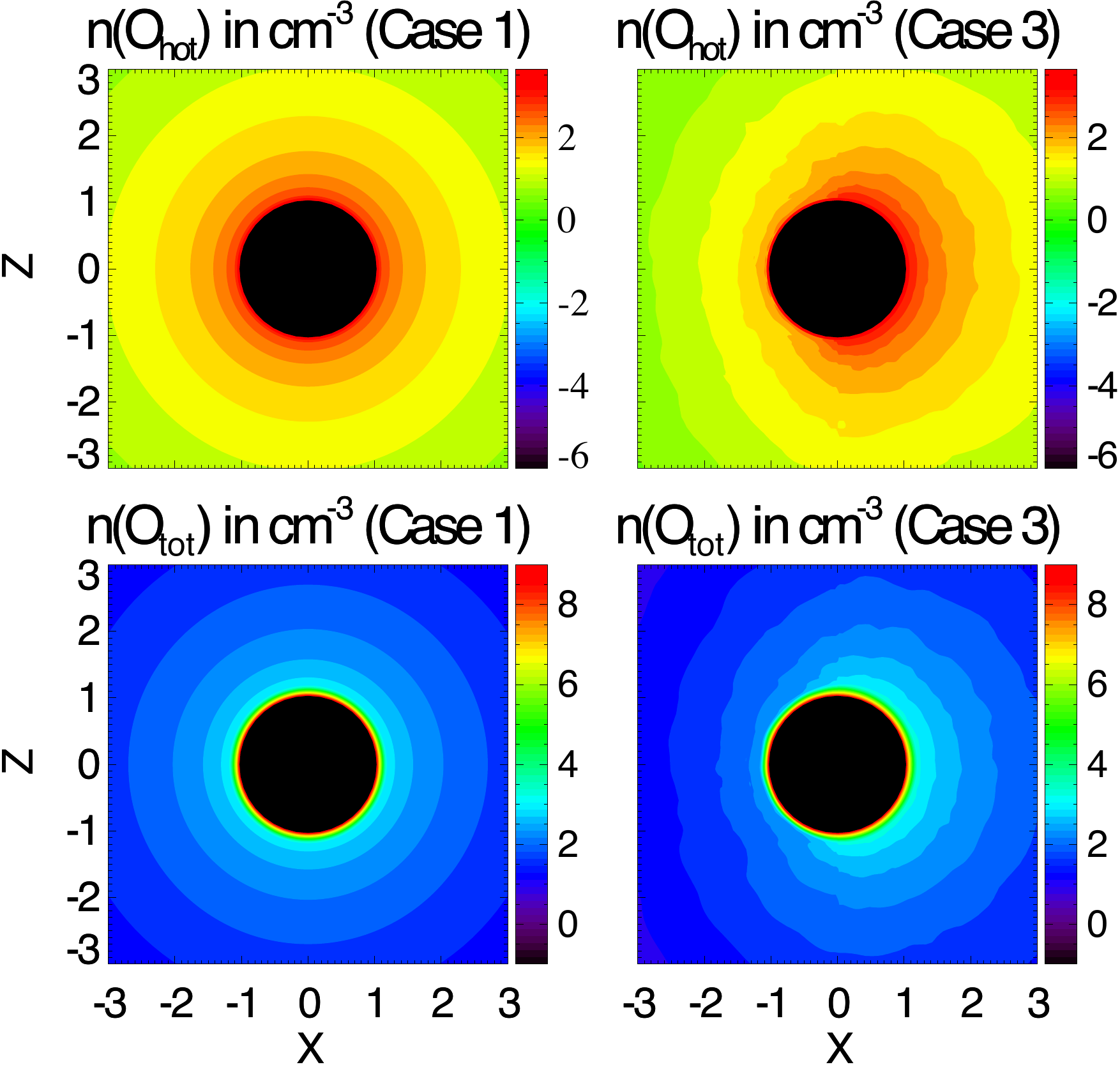}
\caption{Comparisons of the hot atomic oxygen (first row) and total atomic oxygen (the sum of thermal and hot component, second row) density distribution (in cm$^{-3}$) from the globally averaged 1D spherically symmetric AMPS profile (left) and the 3D profile (right) in the x-z meridian plane in the MSO coordinate system. Both cases are based on the aphelion solar moderate conditions (APHMOD). Note the use of different logarithmic scales.}
\label{M-AMPS}
\end{figure*}

Figure \ref{M-AMPS} illustrates the hot and total (the sum of thermal and hot components) atomic oxygen distribution around the Mars globe in a logarithmic scale. Figure \ref{M-AMPS} is based on the Mars-centered Solar Orbital (MSO) coordinate system, where the $+X$ axis points from Mars to the Sun, the $+Z$ axis is perpendicular to the Martian orbital plane and points northward, and the $Y$ axis completes the right-hand system. The left panels correspond to the global hot and total O distribution of the 1D spherically symmetric case, i.e., by averaging over all the longitudes and latitudes from the 3D AMPS output, $\overline{n}(r)=\frac{\oiint n(r,\theta,\phi)\sin\theta d\theta d\phi}{\oiint \sin\theta d\theta d\phi}$. The right panels show the original 3D AMPS and M-GITM output. An inspection of the second row of Figure \ref{M-AMPS} reveals that the thermal atomic oxygen dominates over the hot component at relatively low altitudes (i.e., in the thermosphere), while the hot atomic oxygen is the dominant neutral species at relatively high altitudes (i.e., in the exosphere). Both panels are for the aphelion solar moderate conditions (APHMOD). The 3D AMPS hot oxygen corona shows a great day-night asymmetry which cannot be captured by a 1D spherically symmetric profile. The remarkable asymmetry shown in Figure \ref{M-AMPS} indicates the significance of adopting the 3D hot oxygen corona in a global plasma code for studying the Martian atmospheric ion loss. 

\subsection{BATS-R-US Mars multifluid MHD (MF-MHD) Model}\label{MF-MHDintro}

The 3D BATS-R-US multifluid MHD (MF-MHD) model solves separate continuity, momentum and energy equations for each fluid \citep{Powell1999,Glocer2009,Najib2011,Toth2012,Huang2016,Dong2017}. For the Mars version, it solves MHD equations for four ion fluids H$^+$, O$^+$, O$_2^+$, CO$_2^+$ \citep{Najib2011,Dong2014a,Dong2015a}. Interestingly, \citet{Rubin2014} showed that by using a multifluid MHD model, it can mimic some major features obtained with the hybrid-PIC calculation for a weak comet, such as the finite gyration effect of the planetary/cometary heavy ions and the associated pickup processes. The underlying reason is that MF-MHD includes the dynamics of individual ion species. The Lorentz force term, $\propto (\mathbf{u_s}-\mathbf{u_+}) \times \mathbf{B}$, in the individual ion momentum equation is mainly responsible for the asymmetric ion escape plume and the associated pickup processes, resulting from the difference between the charge averaged ion velocity, $\mathbf{u_+}$, and the individual fluid velocity, $\mathbf{u_s}$, of species $s$ \citep{Dong2014a}. 

At the MF-MHD model lower boundary (100 km above the Martian surface), the densities of O$^+$, O$_2^+$, CO$_2^+$ satisfy the photochemical equilibrium condition \citep[e.g.][chapters 8 and 13]{Nagy2009}. A reflective inner boundary condition for the velocity $\mathbf{u}$ is used, which leads to an approximately zero velocity at the inner boundary as expected. The plasma temperature is set to be twice the value of the neutral temperature at the inner boundary, where both ions and electrons have roughly the same temperature as neutrals due to collisions. We use the 60 degree harmonic expansion model of \citet{Arkani2001} to describe the crustal magnetic fields at Mars \citep{Acuna1999}. The photochemical reactions in the model include charge exchange, photoionization, electron impact ionization and ion-electron recombination. The electron impact ionization rates are given by \citet{Cravens1987}. The elastic collision frequencies are taken from \citet{Nagy2009}. Table \ref{tableRate} summarizes the chemical reactions and the associated rates for inelastic collisions used in the multifluid MHD calculations.
\begin{table}
\centering
\caption{Chemical reactions and associated rates in Mars multifluid MHD code. The ion-neutral and ion-electron chemical reaction rates are adopted from \citet{Najib2011}, while the photoionization frequencies (at the top of atmosphere for aphelion solar moderate conditions) are adopted from \citet{Bougher2015a} as indicated in Figure \ref{coupling}.}\label{tableRate}
\begin{tabular}{lll}
\hline
\hline
\multicolumn{2}{c} {Chemical Reaction} & Rate (s$^{-1}$) \\
\hline
\multicolumn{3}{c} {Primary Photolysis and Particle Impact } \\
\hline
& CO$_{2}$ + $h\nu$ $\rightarrow$  CO$_2^+$ + $e^{-}$     & 8.37$\times$ 10$^{-7}$ \\
& CO$_{2}$ + $h\nu$ $\rightarrow$  CO + O$^+$ + $e^{-}$     &  7.52$\times$ 10$^{-8}$ \\
& O + $h\nu$ $\rightarrow$  O$^+$ + $e^{-}$     &  1.52$\times$ 10$^{-7}$ \\
& H + $h\nu$ $\rightarrow$  H$^+$ + $e^{-}$     &  5.58 $\times$ 10$^{-8}$ \\
& $e^{-}$ + H $\rightarrow$ $e^{-}$ + H$^+$ + $e^{-}$ & see text \\
& $e^{-}$ + O $\rightarrow$ $e^{-}$ + O$^+$ + $e^{-}$ & see text \\
\hline
\multicolumn{2}{c} {Ion-Neutral Chemistry} & Rate (cm$^{3}$ s$^{-1}$) \\
\hline
&  CO$_{2}^+$ + O $\rightarrow$  O$_2^+$ + CO  & $1.64 \times 10^{-10} $   \\
&  CO$_{2}^+$ + O $\rightarrow$  O$^+$ + CO$_2$  & $9.60 \times 10^{-11} $   \\
&  O$^{+}$ + CO$_2$ $\rightarrow$  O$_2^+$ + CO  & $1.1 \times 10^{-9}$ (800/T$_i$)$^{0.39}$   \\
&  O$^{+}$ + H $\rightarrow$  H$^+$ + O  & $6.4 \times 10^{-10} $   \\
&  H$^{+}$ + O $\rightarrow$  O$^+$ + H  & $5.08 \times 10^{-10} $   \\
\hline
\multicolumn{2}{c} {Ion-Electron Recombination Chemistry} & Rate (cm$^{3}$ s$^{-1}$) \\
\hline
&  O$_2^{+}$ + $e^{-}$ $\rightarrow$  O + O  & $7.38 \times 10^{-8}$ (1200/T$_e$)$^{0.56}$   \\
&  CO$_2^{+}$ + $e^{-}$ $\rightarrow$  CO + O  & $3.10 \times 10^{-7}$ (300/T$_e$)$^{0.5}$  \\
\hline
\hline
\end{tabular}
\end{table}

The smallest radial resolution is about 5 km at the inner boundary while the grid size can increase to several thousand kilometers at the outer boundary ($\sim$ 30 R$_M$) due to the nonuniformity in the mesh design. The angular resolution varies from $1.5^{\circ}$ to $3.0^{\circ}$ in a spherical grid mesh bounded by a cube with $-30 R_M \leq X \leq 8 R_M$; $-30 R_M \leq Y,Z \leq 30 R_M$.

\section{Simulation Results and Discussion}

In this section, we discuss the simulation results obtained by using the one-way coupling approach, i.e., both the M-GITM and AMPS neutral profiles are used as the inputs for the MF-MHD model (Figure \ref{coupling}). Firstly, in order to study the effect of the 3D thermosphere on the Martian ionospheric structure and ion escape rates, we adopt either the 1D globally averaged (and thus spherically symmetric) thermosphere (Case 1) or the 3D M-GITM thermosphere (Case 2) while fixing the 1D globally averaged hot oxygen corona. Detailed data-model comparisons along a selected MAVEN trajectory on December 14, 2015 (orbit O2349) are studied. As an illustrative example, we also present the global ionospheric ion distribution at a constant altitude, 200 km, for both Cases 1 and 2. In Section \ref{exosphere}, we investigate the role of the 3D exosphere on the ion escape rate. Three cases are studied for the aphelion solar moderate conditions (APHMOD) with a 1D corona, a 3D corona, and a case without a hot oxygen corona (Cases 2-4). Tables \ref{table1}-\ref{table2} summarize the parameters used for each case. The ion escape rates are summarized in Table \ref{table3}.

\begin{table*}
\caption{Input parameters used for different cases. The solar cycle conditions are chosen based upon one MAVEN trajectory (orbit O2349) on 2015-12-14, during which it has a dayside periapsis.}
\centering
\begin{tabular}{c|c|c|c}
\hline
Simulation \#   & Subsolar Position of Periapsis  & Neutral Atmosphere  & Solar Cycle Conditions  \\
\hline
Case 1  &   167.9$^{\circ}$E, 24.9$^{\circ}$N & 1D$_{cold}$ and 1D$_{hot}$  &  Aphelion Solar  \\ 
Case 2  &   167.9$^{\circ}$E, 24.9$^{\circ}$N & 3D$_{cold}$ and 1D$_{hot}$  &  Moderate  \\
Case 3  &   167.9$^{\circ}$E, 24.9$^{\circ}$N & 3D$_{cold}$ and 3D$_{hot}$  &  (APHMOD)   \\ 
Case 4  &   167.9$^{\circ}$E, 24.9$^{\circ}$N & 3D$_{cold}$ without O$_{hot}$  &   \\ \cline{1-4} 
\hline
\end{tabular}\label{table1}
\end{table*}

\begin{table*}
\caption{Solar wind input parameters used for different cases. The solar wind inputs are taken from MAVEN measurements on 2015-12-14 (orbit O2349), during which it has a steady solar wind and IMF.}
\centering
\begin{tabular}{c|c|c|c|c}
\hline
Simulation \#   & n$_{sw}$ (cm$^{-3}$)   & v$_{sw}$ (km/s)  & IMF (nT) & T$_{sw}^{proton}$ \& T$_{sw}^{electron}$ (K) \\
\hline
Cases 1-4 &   4.85 & (-348.5, -7.0, -25.5)    &  (-0.25, 5.5, -1.0) & 5.9$\times$10$^4$ \& 1.3$\times$10$^5$  \\  
\hline 
\end{tabular}\label{table2}
\end{table*}

\begin{table*}
\centering
\caption{Calculated ion escape rates (in $\times10^{24}$ s$^{-1}$) and molecular to atomic escaping ion ratio listed in the last column.}
\begin{tabular}{c|c|c|c|c|c}
\hline
Simulation cases & O$^+$ & O$_2^+$ & CO$_2^+$ & Total & (O$_2^+$ + CO$_2^+$)/O$^+$ \\ 
\hline
Case 1 (1D$_{cold}$ and 1D$_{hot}$) & 0.57  & 1.45  & 0.29  & 2.30 & 3.06 \\ 
Case 2 (3D$_{cold}$ and 1D$_{hot}$) & 0.74  & 1.27  & 0.27  & 2.28 & 2.09 \\ 
Case 3 (3D$_{cold}$ and 3D$_{hot}$) & 0.89  & 1.18  & 0.31  & 2.38 & 1.67 \\ 
Case 4 (3D$_{cold}$ and noO$_{hot}$) & 0.88  & 1.52  & 0.40  & 2.80 & 2.17 \\ 
\hline
\end{tabular}\label{table3}
\end{table*}

\subsection{Effects of 3D Thermosphere on the Solar Wind-Mars Interaction} \label{thermosphere}

\begin{figure*}[!htb]
\centering
\includegraphics[width=30pc]{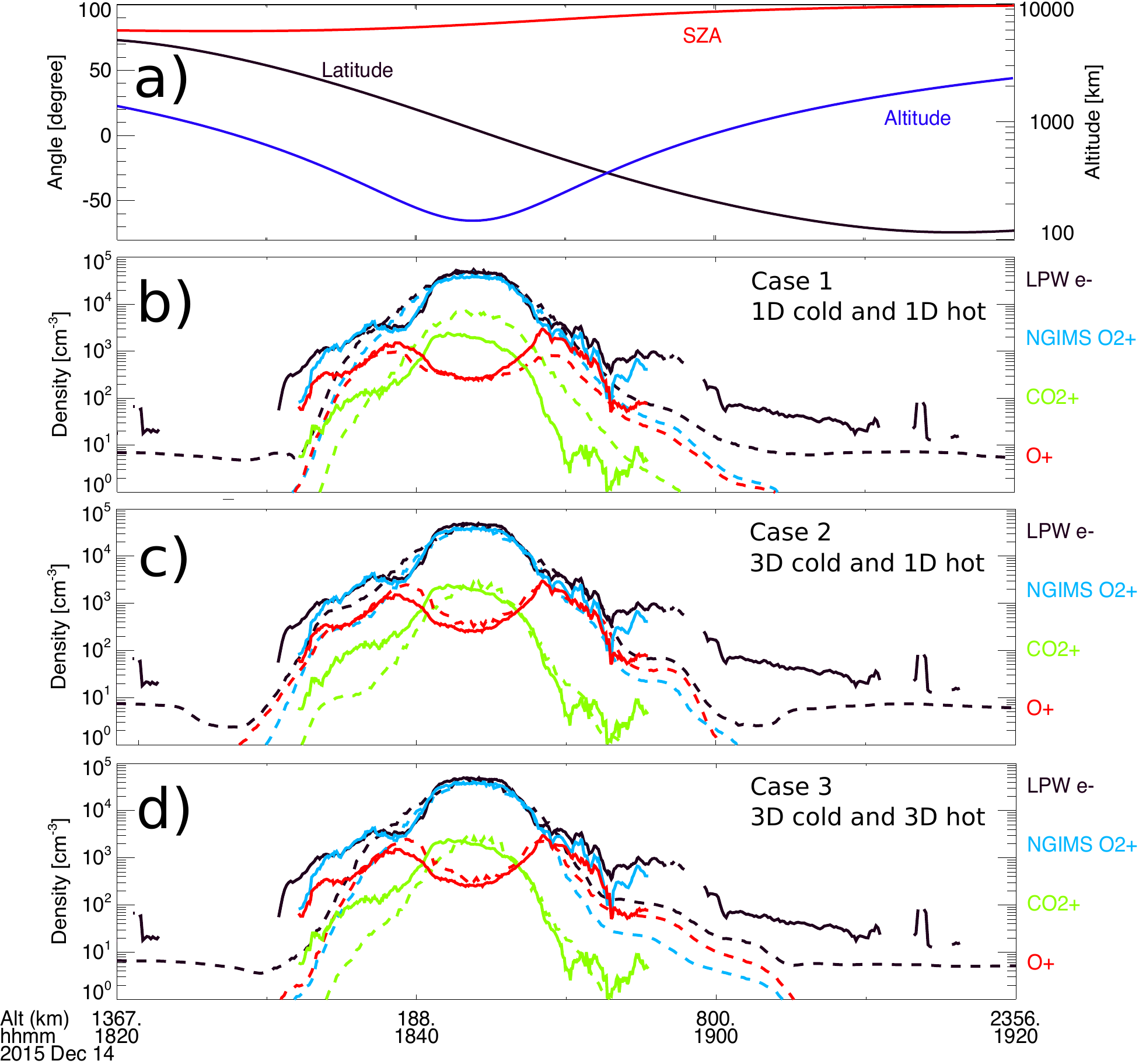}
\caption{Comparisons of the ion and electron densities between the MF-MHD simulations (dashed lines) and the MAVEN observations (solid lines). The ion and electron densities are measured by NGIMS and LPW, respectively. Second panel: Case 1 with 1D thermosphere and exosphere. Third panel: Case 2 with 3D thermosphere and 1D exosphere. Fourth panel: Case 3 with 3D thermosphere and exosphere.}
\label{MAVENc}
\end{figure*}

We first focus on the effect of the 3D thermosphere on the solar wind-Mars interaction. Figure \ref{MAVENc} presents the data-model comparison of the ionospheric density profiles between the MF-MHD calculations (dashed lines) and the MAVEN data (solid lines). The electron density was measured by the Langmuir Probe and Waves (LPW) instrument on board the spacecraft. The O$^+$, O$_2^+$ and CO$_2^+$ ion densities were measured by the Neutral Gas and Ion Mass Spectrometer (NGIMS). Figure \ref{MAVENc}a depicts the spacecraft altitudes (blue), latitude (black) and solar zenith angle (red) versus time along the trajectory. The rest of the panels are the detailed data-model comparisons based on the 1D thermosphere and exosphere (Case 1, Figure \ref{MAVENc}b), the 3D thermosphere and the 1D exosphere (Case 2, Figure \ref{MAVENc}c), and the 3D thermosphere and exosphere (Case 3, Figure \ref{MAVENc}d), respectively. In Figure \ref{MAVENc}, the MF-MHD model displays the maximum ionospheric ion and electron densities at the periapsis of orbit O2349 (on December 14, 2015), in good agreement with the MAVEN observation. Both the MF-MHD calculations and the MAVEN data reveal that O$_2^+$ is the dominant ion in the Martian ionosphere.

\begin{figure*}[!htb]
\centering
\includegraphics[width=30pc]{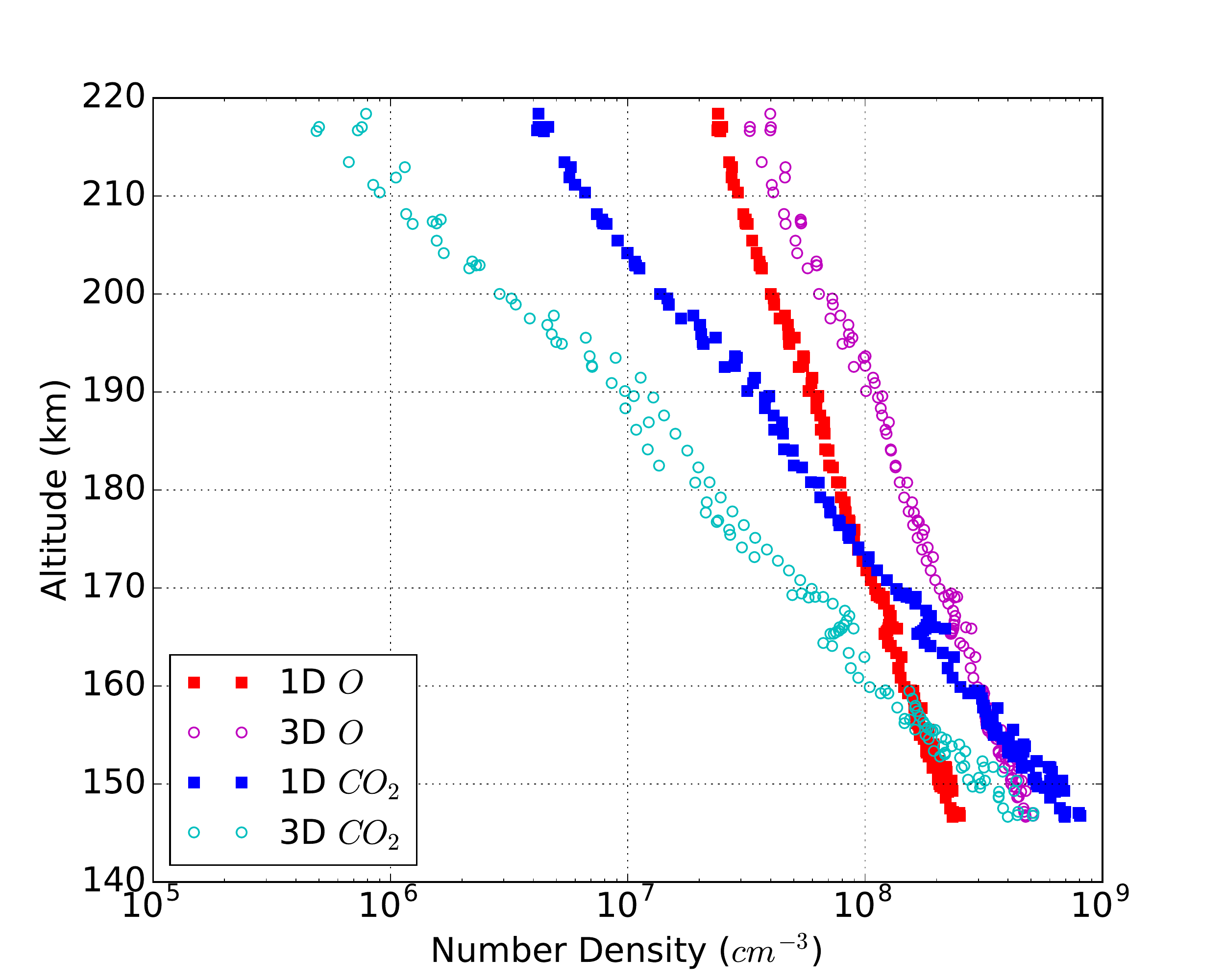}
\caption{Comparisons between the 1D and 3D thermospheric density profiles along the MAVEN trajectory. The plot includes both inbound and outbound data.}
\label{1D3DhotO}
\end{figure*}

Compared with Case 1, the MF-MHD calculations based on the 3D M-GITM thermosphere (Case 2) fit the observational data better, demonstrating the importance of adopting the 3D thermosphere in a global plasma code. In Figure \ref{MAVENc}b, the calculated molecular ion (O$_2^+$ and CO$_2^+$) densities along the MAVEN trajectory are slightly higher than the NGIMS data whilst the O$^+$ ion density is slightly lower than that observed. Figure \ref{MAVENc}c, however, shows an opposite trend as presented in Figure \ref{MAVENc}b. In order to understand the deviation between simulations and observations, we plot both 1D and 3D thermospheric O and CO$_2$ densities along the MAVEN trajectory (Figure \ref{1D3DhotO}). As we expected, the 3D thermosphere (Case 2) has a higher O and lower CO$_2$ abundance compared to the 1D thermosphere (Case 1) along the MAVEN trajectory. This helps to explain the variation trend in the ion densities from Figure \ref{MAVENc}b to Figure \ref{MAVENc}c. An inspection of Figure \ref{MAVENc}c and Figure \ref{MAVENc}d reveals that the 3D hot oxygen does not have a significant effect on the ionospheric density distribution compared to the 1D exosphere case. 

\begin{figure*}[!htb]
\centering
\includegraphics[width=33pc]{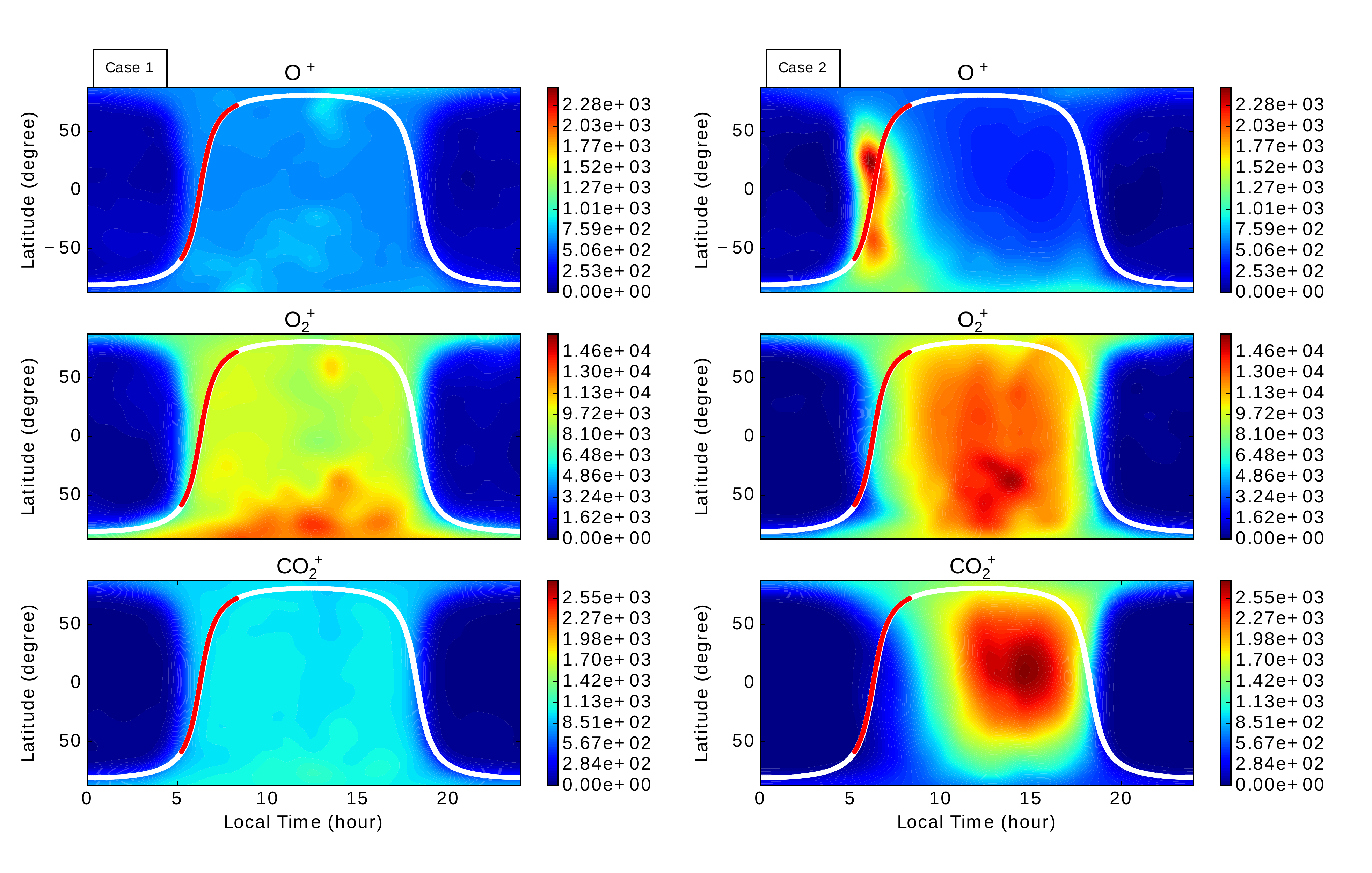}
\caption{The ionospheric density maps for O$^+$, O$_2^+$, and CO$_2^+$ at 200-km altitude for Case 1 (left column) and Case 2 (right column). The thick white curve in each panel represents the projection of a selected MAVEN trajectory (orbit O2349 on December 14, 2015). The red segment corresponds to the regions with altitudes lower than 1000 km, including the periapsis. Note the use of different colorbar range in different rows.}
\label{2Diono}
\end{figure*}

Although we have presented the ionospheric ion densities along one MAVEN trajectory, it is also important to depict the global density distribution of the Martian ionosphere. Figures \ref{2Diono} illustrates the 2D (latitude \emph{vs.} local time, at 200-km altitude) ionospheric maps from the MF-MHD model for Case 1 (left column) and Case 2 (right column). For both cases, the top panels show the density distribution of O$^+$, and the middle to bottom panels display the density distribution of O$_2^+$ and CO$_2^+$, respectively. The smooth transition of the ion density around terminator region is a result of the implementation of Chapman function from \citet{Smith1972} in the MF-MHD model. Again, both columns show that O$_2^+$ is the dominant ion species in the Martian ionosphere. In order to better understand the different ion distributions shown in Figure \ref{MAVENc}, we also plotted the projection of the MAVEN trajectory in each panel.

\begin{figure*}[!htb]
\centering
\includegraphics[width=25pc]{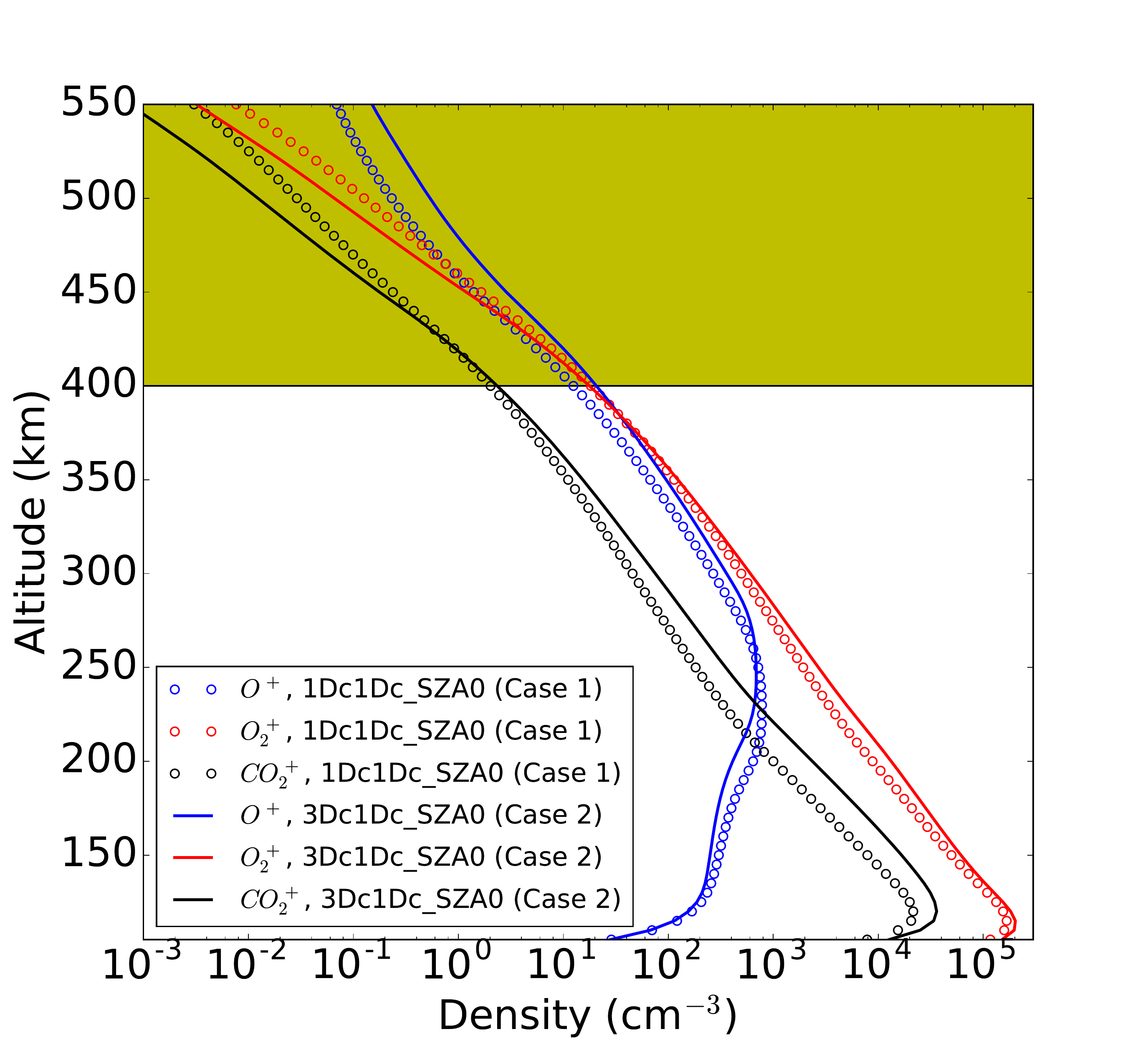}
\caption{The vertical ionospheric density profiles for O$^+$, O$_2^+$, and CO$_2^+$ at SZA=0 for Case 1 (circle markers) and Case 2 (solid curves). The yellow shading highlights a region that is more relevant to the ion escape.}
\label{1Diono}
\end{figure*}

In Figure \ref{2Diono} (left column), all the ions mirror a similar ionospheric pattern as a result of the 1D spherically symmetric thermospheric input. The enhanced ion density in the southern hemisphere is mainly caused by the crustal magnetic fields given that the crustal anomalies are shifted to higher solar zenith angles (i.e. the southern polar region in MSO) at aphelion. The same enhancement at the southernmost latitudes seen in the left column is not present in the right column because the 3D asymmetric thermosphere (as shown in Figure \ref{MGITM}) produces relatively high ion abundance at lower latitudes and northern hemisphere compared to the 1D thermosphere case. In Figure \ref{2Diono} (right column), however, the ionospheric global distributions between molecular ions (O$_2^+$ and CO$_2^+$) and atomic ions (O$^+$) are distinct when adopting the 3D thermosphere. On the other hand, O$_2^+$ and CO$_2^+$ share similar ionospheric patterns. It is well known that the Martian dayside ionosphere is triggered by the photoionization resulting from the solar EUV radiation \citep{Bougher2008}. Subsequently, the photoionized CO$_2^+$ quickly reacts with neutral O to produce the major ionospheric species O$_2^+$; therefore, O$_2^+$ exhibits a similar ionospheric distribution as CO$_2^+$. The ionospheric density peaks of O$_2^+$ and CO$_2^+$ are also located at almost the identical altitude (e.g., see Figure \ref{1Diono}). Although thermospheric O can be photoionized by photons (the main channel), ionized through charge exchange with other ion species, and impact ionized by electrons to produce O$^+$, the absence of the neutral oxygen atom (in the dayside thermosphere in 3D case) leads to a low abundance of O$^+$ in the dayside ionosphere (right column of Figures \ref{2Diono}), consistent with the neutral density distribution shown in Figure \ref{MGITM}. Compared with the previous work, O$^+$ in the present work can also be produced through photoionization of CO$_2$ as a secondary channel.

In order to understand the effect of the 3D thermosphere on the ion escape, we calculate the ion escape rates and list them in Table \ref{table3}. The calculations are conducted by integrals of the plasma density multiplied by the radial velocity component at the surface of a sphere far from the planet. Calculations (not presented here for the sake of brevity) show that ion escape rates do not change to any significant degree ($< \sim$5\%) once the radius exceeds 5 R$_M$, the result presented in the remainder of this paper use the integral sphere with radius 6 R$_M$. Compared with Case 1 (with 1D globally averaged thermosphere), the O$^+$ ion escape rate in Case 2 (with 3D thermosphere) increases whilst molecular ionospheric ion (O$_2^+$ and CO$_2^+$) escape rate deceases. These trends can be explained by the vertical ionospheric density profiles (at SZA=0) shown in Figure \ref{1Diono}. As seen from Figure \ref{1Diono}, more O$^+$ at lower altitudes for Case 1 and more molecular ionospheric ions (O$_2^+$ and CO$_2^+$) at lower altitudes for Case 2, consistent with Figure \ref{2Diono}. The relative abundance, however, shows a contrary trend at high altitudes (in the yellow shading region). Interestingly, the high-altitude ion abundance is consistent with the ion escape rates listed in Table \ref{table3} since only those ions above a certain altitude (i.e., ion exobase) are able to escape \citep[e.g.,][]{Cravens2017}. From test-particle simulations, \citet{Fang2010b} also found that generally on the dayside, only less than 35\% of ions are able to escape below 400-km altitude.

\subsection{Effects of 3D Exosphere on the Solar Wind-Mars Interaction} \label{exosphere}

In order to investigate the effect of the 3D exosphere (i.e., hot oxygen corona) on the interplay between the Martian upper atmosphere and the solar wind, we study three cases with a 1D corona, a 3D corona, and a case without a hot oxygen corona (Cases 2-4). This is similar to the study by \citet{Curry2013b}, who conducted test-particle simulations to study O$^+$ ion loss rates with and without the 1D hot oxygen corona from \citet{Kim1998} by fixing the 1D thermosphere from \citet{Ma2004}.

\subsubsection{Effects of 3D Hot Oxygen Corona on O$^+$ Ion Escape}

\begin{figure*}[!htb]
\centering
\includegraphics[width=13pc,angle=90]{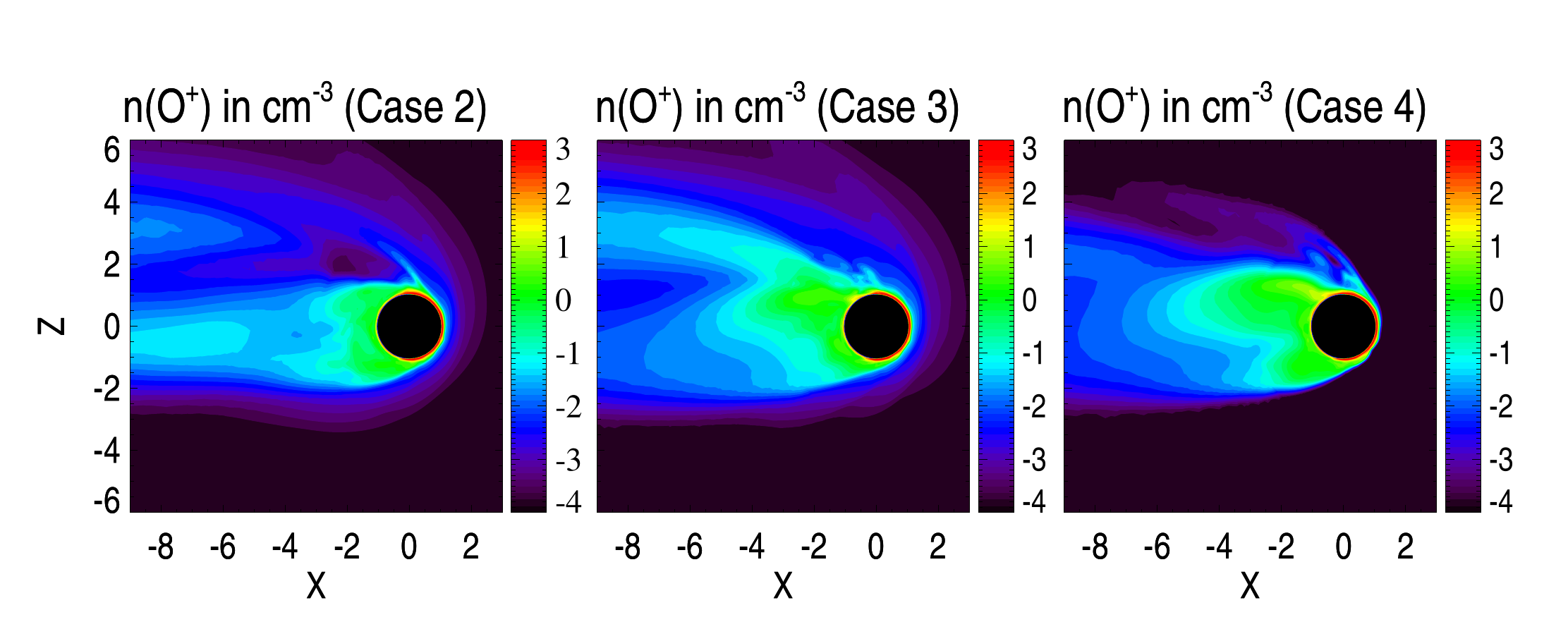}
\caption{Comparisons of O$^+$ density in the $x$-$z$ plane for a 1D corona, a 3D corona, and a case without a hot oxygen corona. Left: case with the 1D globally averaged hot oxygen corona (Case 2). Middle: case with the 3D AMPS hot oxygen corona (Case 3). Right: case without the hot oxygen corona (Case 4). Note the use of a logarithmic scale.}
\label{mhdhotO}
\end{figure*}

\begin{figure*}[!htb]
\centering
\includegraphics[width=25pc,angle=0]{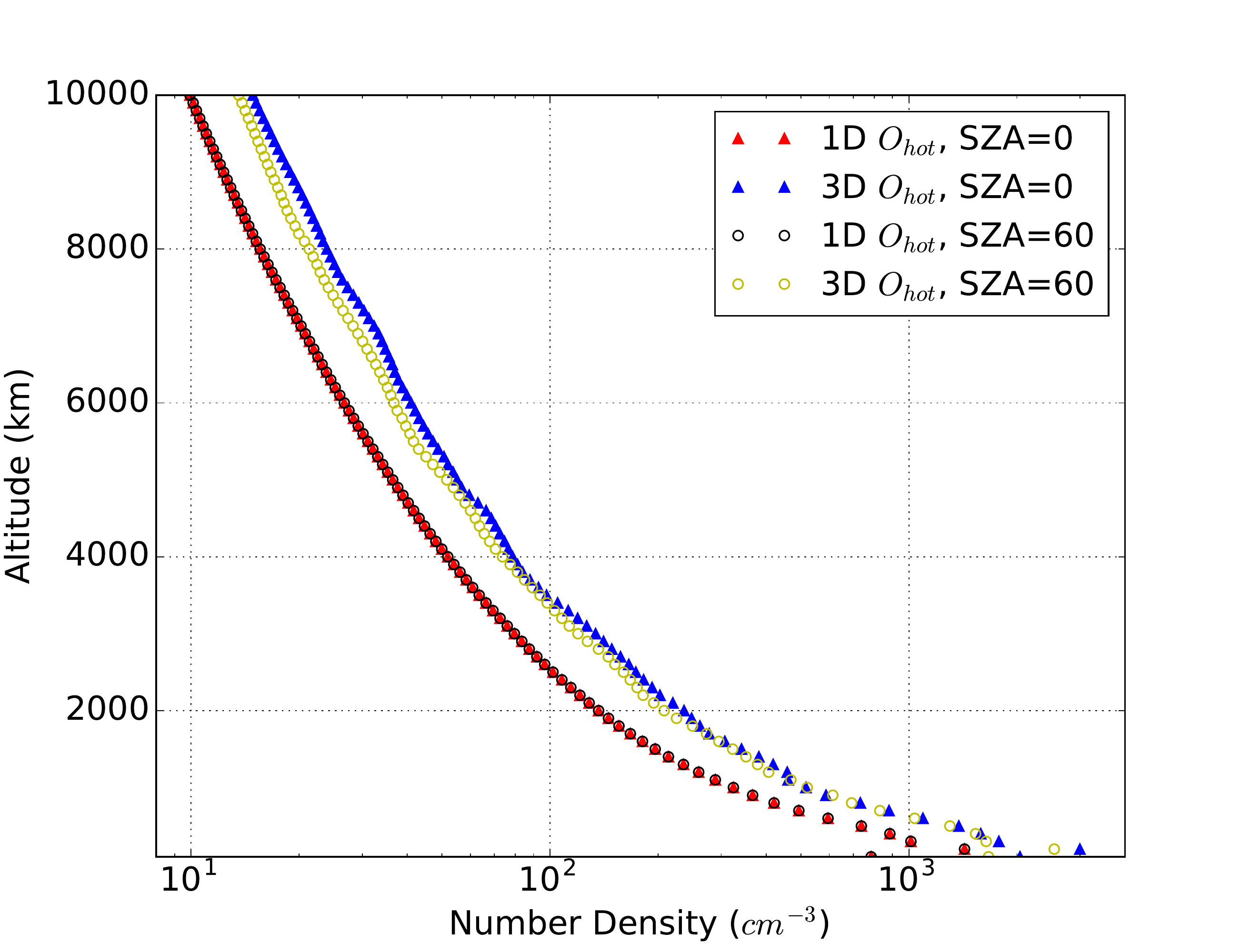}
\caption{The comparison between 1D and 3D exospheric hot oxygen profiles at SZA=0 and SZA=60.}
\label{1DhotO}
\end{figure*}

Figure \ref{mhdhotO} depicts the O$^+$ density in the $x$-$z$ plane. One of the features of the MF-MHD model is that it can capture the asymmetric escape plume of the planetary pickup ions \citep{Najib2011,Dong2014a,Rubin2014}. Both Cases 2 and 3 present dayside pickup O$^+$ ion escaping from the extended hot oxygen corona region whilst the high-altitude corona O$^+$ ions cannot be observed in Case 4 due to the absence of an atomic oxygen source. Compared with Case 2, more O$^+$ are present in the dayside exospheric region in Case 3, consistent with those hot oxygen density distributions shown in Figure \ref{1DhotO}. All three cases present a large number of O$^+$ ions escaping from the nightside plasma wake region as well. The color contours in Figure \ref{mhdhotO} can be used to explain why the O$^+$ ion escape rate of Case 2 is smallest among three cases and O$^+$ ion escape rates between Case 3 and Case 4 are similar. The similar ion escape rate between Case 3 and Case 4 implies that the thermospheric oxygen atoms make a significant contribution on the O$^+$ ion escape rate.

\subsubsection{Effects of 3D Hot Oxygen Corona on Ion Escape: O$^+$ \emph{vs.} (O$_2^+$ and CO$_2^+$)}

In this section, we will focus on Cases 3-4 and aim to understand the effect of a 3D hot oxygen on O$^+$ ion escape versus ionospheric molecular ion (O$_2^+$ and CO$_2^+$) losses. We summarize the calculated ion escape rates for Cases 3-4 in Table \ref{table3}.

Compared with Case 3 that includes hot oxygen, Case 4 (without hot oxygen) has higher O$_2^+$ and CO$_2^+$ escape rates but maintains a similar value of O$^+$ escape rate. The variation in the molecular to atomic escaping ion ratio (i.e., the last column of Table \ref{table3}) indicates that the hot oxygen component has a shielding effect that can protect the Martian ionosphere from the solar wind erosion, especially for O$_2^+$ and CO$_2^+$ that have a relatively high mass and thus are located at relatively low altitudes. It also reveals that the thermospheric oxygen is the dominant neutral source in determining the Martian O$^+$ ion escape for APHMOD under the nominal solar wind conditions.

Before proceeding further, recall that  \citet{Dong2015a} calculated the ion escape rates under different solar cycle and seasonal conditions. They found that O$_2^+$ is the dominant escaping ion at solar minimum whilst O$^+$ is the dominant escaping species at solar maximum. \citet{Curry2013} also showed the importance of O$^+$ ion escape using a test-particle model at solar maximum. Therefore, we conduct a case study by choosing Case 10 (autumnal equinox solar maximum - AEQUMAX) in \citet{Dong2015a} with and without 3D hot oxygen corona; the O$^+$ ion escape rates are 4.57$\times$10$^{24}$ s$^{-1}$ and 3.70 $\times$10$^{24}$ s$^{-1}$, respectively. Compared with APHMOD, the hot oxygen becomes more important for O$^+$ ion escape at AEQUMAX. The thermospheric oxygen atom, however, is still the primary neutral source for O$^+$ ion escape for AEQUMAX under the nominal solar wind conditions.

The keys to understand the shielding effect of the hot oxygen corona are the ion pickup and mass loading processes. Given the momentum and energy conservation, the solar wind momentum and energy fluxes start to gradually decrease when approaching Mars due to the mass loading of high-altitude O$^+$ (ionized from hot oxygen corona). In the absence of a hot oxygen corona, the solar wind can directly interact with the Martian ionosphere and thermosphere.  

\begin{figure*}[!htb]
\centering
\includegraphics[width=34pc]{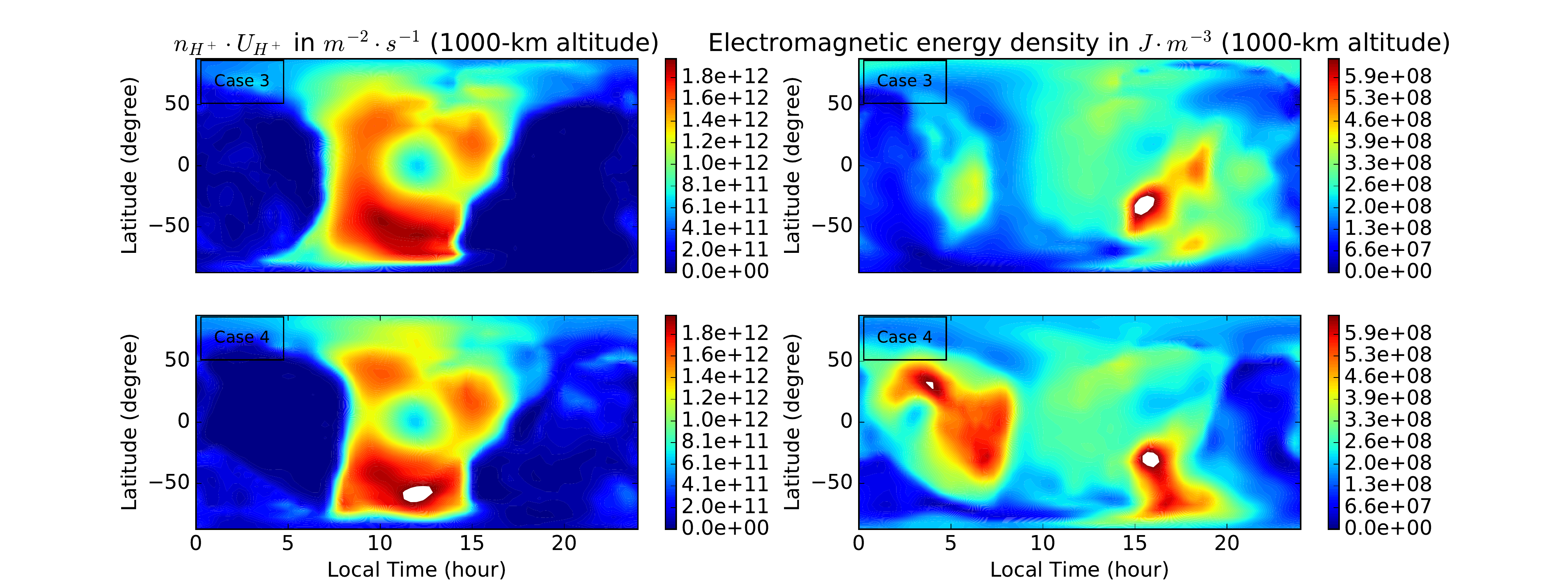}
\caption{Comparisons of the solar wind proton number flux, $n_{H^+}U_{H^+}$ between Case 3 and Case 4 (left). Comparisons of the electromagnetic energy density, $\mathcal{E}$ between Case 3 and Case 4 (right). Both are depicted at 1000-km altitude above the Martian surface. The white color highlights the regions beyond the high saturation of the colorbar.}
\label{ionospeed}
\end{figure*}

Figure \ref{ionospeed} demonstrates the solar wind proton number flux (left) and the electromagnetic energy density (right) at 1000 km above the Martian surface for Cases 3 and 4. The electromagnetic energy density is defined as 

\begin{equation}
\mathcal{E} = \frac{\epsilon_0E^2}{2}  + \frac{B^2}{2\mu_0}
\end{equation}
where $\epsilon_0$ and $\mu_0$ are the permittivity and permeability of free space, respectively. $E$ denotes the electric field (see Eq.(2) in \citet{Dong2014a}) and $B$ represents the magnetic field. In Figure \ref{ionospeed} (left panel), the proton number flux in Case 4 is saturated at 1000-km altitude (in white) while no saturation is observed in Case 3 using the same colorbar range. Compared with Case 4, less electromagnetic energy density (right panel of Figure \ref{ionospeed}) is available at same altitude in Case 3, indicating the ionosphere is more disturbed by the solar wind without a hot oxygen corona.


\section{Conclusions}

Recently, \citet{Dong2017b,Dong2018} studied the atmospheric ion escape of exoplanets (such as Proxima b and the TRAPPIST-1 system by assuming Venus-like atmospheres) orbiting M-dwarfs in the close-in habitable zone. Due to the strong EUV flux and extreme stellar wind parameters, they found that the O$^+$ ion is always the dominant escaping ion species (due to its relatively small mass and thus large scale height) compared to O$_2^+$ and CO$_2^+$. In certain circumstances, the ionospheric molecular ion (O$_2^+$ and CO$_2^+$) escape rates of Venus-like exoplanets orbiting M-dwarfs are similar to (and even smaller than) the cases in our solar system despite the much more intensive stellar radiation and stellar wind, as a result of the short star-planet distance, e.g, 0.05 AU for Proxima b. The underlying reason is that the mass loading of relatively light O$^+$ ion slows down the stellar wind. At ancient times, the EUV flux and solar wind parameters were much stronger than that of the current epoch (partly resembling those of the M-dwarf exoplanets discussed earlier), and Mars also has a much more extensive and intensive hot oxygen corona \citep{Valeille2010}, indicating that hot oxygen exosphere may provide an important source for O$^+$ ion escaping billions of years ago. Therefore, the hot oxygen corona may play a crucial role in the long-term evolution of the Martian atmosphere and its composition over its history \citep{Dong2014b,Dong2018b}. Based on this study, we speculate that the early loss rate of the ionospheric molecular ions (O$_2^+$ and CO$_2^+$) may be even lower than the current value due to the strong shielding (i.e., mass loading) effect of high-altitude oxygen ions. 

In summary, we studied the solar wind interaction with the Martian upper atmosphere using a one-way coupled framework of three comprehensive 3D models, i.e., the M-GITM thermosphere output and the Mars AMPS hot atomic oxygen corona are used as the inputs for the MF-MHD model. The effects of 1D and 3D \emph{cold} thermosphere and \emph{hot} oxygen corona on the ionospheric structure and ion escape rates are studied in detail by comparing four selected cases. While the total ion escape rates by adopting 1D and 3D neutral atmospheres are similar, the detailed ionospheric density distributions are distinguishable. Compared with the 1D thermosphere, the MF-MHD calculations based on 3D thermosphere are in better agreement with MAVEN observations. We also found that the hot oxygen corona plays an important role in protecting the Martian ionosphere and thermosphere from the solar wind erosion, i.e., reducing the molecular ionospheric ion (O$_2^+$ and CO$_2^+$) escape rate. The shielding effect can be explained by the mass loading of the high-altitude hot oxygen ions. Moreover, the simulation results reveal that the \emph{cold} oxygen is the primary neutral source for O$^+$ ion escape during this unusually quiet solar cycle. 

\acknowledgments
The authors thank M. Lingam for the helpful discussions and comments. This research was supported by NASA grant NNH10CC04C through MAVEN project, and NASA grants 80NSSC18K0288, NNX14AH19G and NNX16AQ04G. Resources supporting this work were provided by the NASA High-End Computing (HEC) Program through the NASA Advanced Supercomputing (NAS) Division at Ames Research Center. We also would like to acknowledge high-performance computing support from Yellowstone (ark:/85065/d7wd3xhc) and Cheyenne (doi:10.5065/D6RX99HX) provided by NCAR's Computational and Information Systems Laboratory, sponsored by the National Science Foundation. The Space Weather Modeling Framework that contains the BATS-R-US code used in this study is publicly available from \texttt{http://csem.engin.umich.edu/tools/swmf}. The MAVEN data is publicly available through the Planetary Plasma Interactions Node of the Planetary Data System \texttt{https://pds-ppi.igpp.ucla.edu/mission/MAVEN}. The model results are publicly available at \texttt{https://umich.box.com/s/tas6l27xzupb4dvkp2whz7k6wklnwes3}.






\bibliographystyle{agufull08}
\listofchanges

\end{document}